\documentclass[twocolumn]{wiley-article}
\usepackage{natbib}
\setcitestyle{square,super,compress,comma}

\usepackage{siunitx}
\let\vec\relax
\DeclareMathAccent{\vec}{\mathord}{letters}{"7E}
\def\citep#1{(\cite{#1})}
\usepackage{mhequ}
\usepackage{graphicx}
\usepackage{amsmath}
\usepackage{amssymb}
\usepackage{breakcites}
\usepackage{wrapfig}
\usepackage{xcolor}

\usepackage[utf8]{inputenc} 
\usepackage[T1]{fontenc}    
\usepackage[hidelinks]{hyperref}       
\hypersetup{
    colorlinks = true,
    linkcolor = [rgb]{0,0.2,0.65},
    filecolor = black,      
    urlcolor = blue,
    citecolor = [rgb]{0,0.55,0.55}
    }

\usepackage{url}            
\usepackage{booktabs}       
\usepackage{amsfonts}       
\usepackage{nicefrac}       
\usepackage{microtype}      
\usepackage{graphicx}
\graphicspath{ {./images/} }

\usepackage{fullwidth}

\usepackage{newfloat}
\DeclareFloatingEnvironment[fileext=lop]{FigS}
\usepackage[labelfont=bf]{caption}

\usepackage{import}

\usepackage{setspace}
\usepackage{siunitx}
\usepackage{times}
\usepackage{lettrine}
\usepackage[skip=0.3cm]{parskip}

\def\l{\left}
\def\r{\right}
\def\mut{\mathrm{mut}}
\def\Ber{\mathrm{Ber}}
\def\Pr{\mathrm{Pr}}
\def\relfit{\textrm{relfit}}
\def\cumfit{\textrm{cumfit}}
\def\allfit{\textrm{allfit}}
\def\minfit{\textrm{minfit}}
\def\Anc{\mathrm{Anc}}
\def\denovo{\textit{de novo}~}
\usepackage{xcolor}
\definecolor{YKB}{rgb}{0.05,0.18,0.60}

\papertype{Hypothesis}
\paperfield{Insights \& Perspectives}

\title{\denovo gene birth as an inevitable consequence of adaptive evolution}

\author[1]{Somya Mani}
\author[1,2]{Tsvi Tlusty}

\affil[1]{Center for Soft and Living Matter, Institute for Basic Science, Ulsan 44919, Republic of Korea}
\affil[2]{Departments of Physics and Chemistry, Ulsan National Institute of Science and Technology (UNIST), Ulsan 44919, Republic of Korea}

\corraddress{~}
\corremail{tsvitlusty@gmail.com}

\fundinginfo{Institute for Basic Science, Grant IBS-R020.}

\runningauthor{Mani and Tlusty}

\begin{document}
\setcitestyle{super} 
\begin{frontmatter}
\maketitle
\begin{abstract}
  Phylostratigraphy suggests that new genes are continually born \denovo from non-genic sequences, and the genes that persist found new lineages, contributing to the adaptive evolution of organisms. While recent evidence supports the view that \denovo gene birth is frequent and widespread, the mechanisms underlying this process are yet to be discovered. Here we hypothesize and examine a potential general mechanism of gene birth driven by the accumulation of beneficial mutations at non-genic loci. To demonstrate this possibility, we model this mechanism within the boundaries set by current knowledge on mutation effects. Estimates from this analysis are in line with observations of recurrent and extensive gene birth in genomics studies. Thus, we propose that, rather than being inactive and silent, non-genic regions are likely to be dynamic storehouses of potential genes.
\keywords{ \denovo gene birth, spontaneous mutation, distribution of fitness effects, adaptation}
\end{abstract}

\end{frontmatter}

\section*{Introduction}
 Broadly, a gene is defined as a sequence in the genome which yields phenotypic traits through regulatory interactions of its products with other genes and the environment \cite{portin2017evolving}. For long, the answer to the question of origin of genes was taken for granted; it was believed that a basic set of `founder genes', numbering some thousands, originated a long time ago, and all new genes are exclusively derived from these founder genes. In contrast to this `genes come from genes' picture, genomic evidence indicates that \SIrange{10}{30}{\percent} of all genes across eukaryotic genomes are orphan genes, for which no homology can be detected with any established conserved genes~\cite{tautz2011evolutionary}. While it is possible that orphan genes are products of duplication and rapid divergence of conserved genes, recent studies indicate that most orphan genes are likely to be generated \denovo, starting from previously non-genic sequences~\cite{vakirlis2020synteny}. Moreover, phylostratigraphy suggests that such new genes are generated continuously~\cite{tautz2011evolutionary}. Much of our current understanding of the rates and extent of \denovo gene birth comes from genomics studies. But to understand the general mechanisms underlying the process, there is a need for independent theoretical models that build upon basic evolutionary processes. In this work,  we propose one such basic mechanism --- \denovo gene birth through the accumulation of beneficial mutations--- and demonstrate it using a simple mathematical model. 

Known \denovo genes display some intriguing patterns: new genes are born preferably in genomic regions with high GC content and near meiotic hot spots. In animals, new genes are more likely to be expressed in the brain and testis~\cite{vakirlis2018molecular}. Interestingly, these cells and genomic regions are also especially prone to pervasive, leaky transcription~\cite{jensen2013dealing}. These observations point to a possible mechanism of gene birth~\cite{van2019novo}--- non-genic loci, made visible to natural selection by pervasive expression, can be driven to evolve adaptively and gain new functions and thereby lead to \denovo gene birth~\cite{carvunis2012proto}.

Two simple studies, taken together, lend support to such a mechanism: first, random sequences can gain functionality, provided they are consistently expressed~\cite{hayashi2003can}. And second, it was demonstrated that new promoters could easily evolve in \textit{E. coli}~\cite{yona2018random}. These studies highlight the possibility that non-genic sequences can, in stages, gain the hallmarks of genes: \textit{regulated expression} and \textit{functionality}.
    
We draw on these observations and further propose that gene birth can be understood as an \textit{inevitable} consequence of adaptive evolution of non-genic sequences. To illustrate this, we present a blueprint for a minimal model of gene birth that uses characteristics of spontaneous mutations, the simplest units of adaptive evolution, as its building blocks. Specifically, we consider the process by which a locus that initially provides very low fitness advantage, corresponding to leaky expression, starts to accumulate beneficial mutations. We conjecture that this process reflects the early stages of \denovo gene birth.  
    

In the model, we define both genes and mutations in a coarse-grained manner, purely in terms of their fitness contributions. Practically, genes can be defined at many levels; genes are currently described as conserved genomic sequences that produce functional products. Function and regulated expression are fundamental hallmarks of a gene, and gene birth is essentially the process of concerted evolution of these aspects. But the dearth of quantitative data that describe the evolution of expression levels and potential functionality of non-genic sequences constrains us to employing an abstract definition of genes. Nevertheless, this simplification allows us to leverage currently available data to produce biologically reasonable estimates. 

Similarly, we describe mutations also in terms of their fitness effects. Experimentally, fitness effects of spontaneous mutations are assessed through mutation accumulation studies. These studies directly~\cite{sane2020shifts}, or indirectly~\cite{bondel2019inferring} allow inference of the fitness effects of single mutations, thereby yielding a distribution of fitness effects (DFE). The key assumption of our model is that the DFE of small loci (\numrange{100}{1000} base pairs) are similar in form to the DFE across the whole genome, which is the quantity measured in mutation accumulation experiments. In particular, we assume that the DFE of loci that start out as non-genic and are evolving adaptively, can be captured using the same parameters that are used to describe the DFE of whole genomes. These assumptions are supported by observations in ~\cite{bondel2019inferring}, that the DFE of specific regions of the genome, such as exons, introns or intergenic sequences, are similar to each other and to the DFE of the whole genome. Now in general, the DFE is known to differ across different regions of the genome~\cite{racimo2014approximation}, and across different species~\cite{huber2017determining}. We accommodate this diversity by sampling a  wide range of DFEs, which differ in the frequency and size of beneficial and deleterious mutations, and also in the shape of the distribution, to assess the conditions under which spontaneous mutations can lead to gene birth. 

Our simple analysis indicates that gene birth should be highly likely for a wide range of DFEs, given a time frame of millions of years. Especially, we find that the presence of rare, large-effect mutations can potentially compensate for beneficial mutations being small and infrequent on average. We also tested the more realistic scenario where the DFE of a genomic locus fluctuates over long periods of time; under these conditions, gene birth becomes virtually inevitable.

Thus, we propose the intriguing hypothesis that \denovo gene birth through the adaptive evolution of non-genic sequences is practically unavoidable. We also discuss experiments to test the consequences of this hypothesis. We anticipate that in the future, experiments would characterize not only the fitness effects of mutations, but also distinguish between their effect on expression level of loci and functionality of expression products. Such data should inform more detailed models that capture the essence of genes, and therefore of the process of \denovo gene birth.


\section*{An adaptive model of gene birth}

We use the Wright-Fisher framework to model well-mixed populations of fixed size $N$, composed of asexually reproducing haploid individuals. For each individual $i$, we consider the fitness contribution $F_i$ of a single locus in its genome. Here, fitness represents exponential growth rate, which is equivalent to the quantities considered in experiments that measured DFEs (e.g.,~\cite{bondel2019inferring}). Since our definition of genes is not tied to any specific function, we describe a locus as \textit{genic} if it consistently contributes a fitness advantage above a predetermined \textit{genic threshold} and non-genic otherwise. 

\begin{figure*}

\includegraphics[width=1\linewidth]{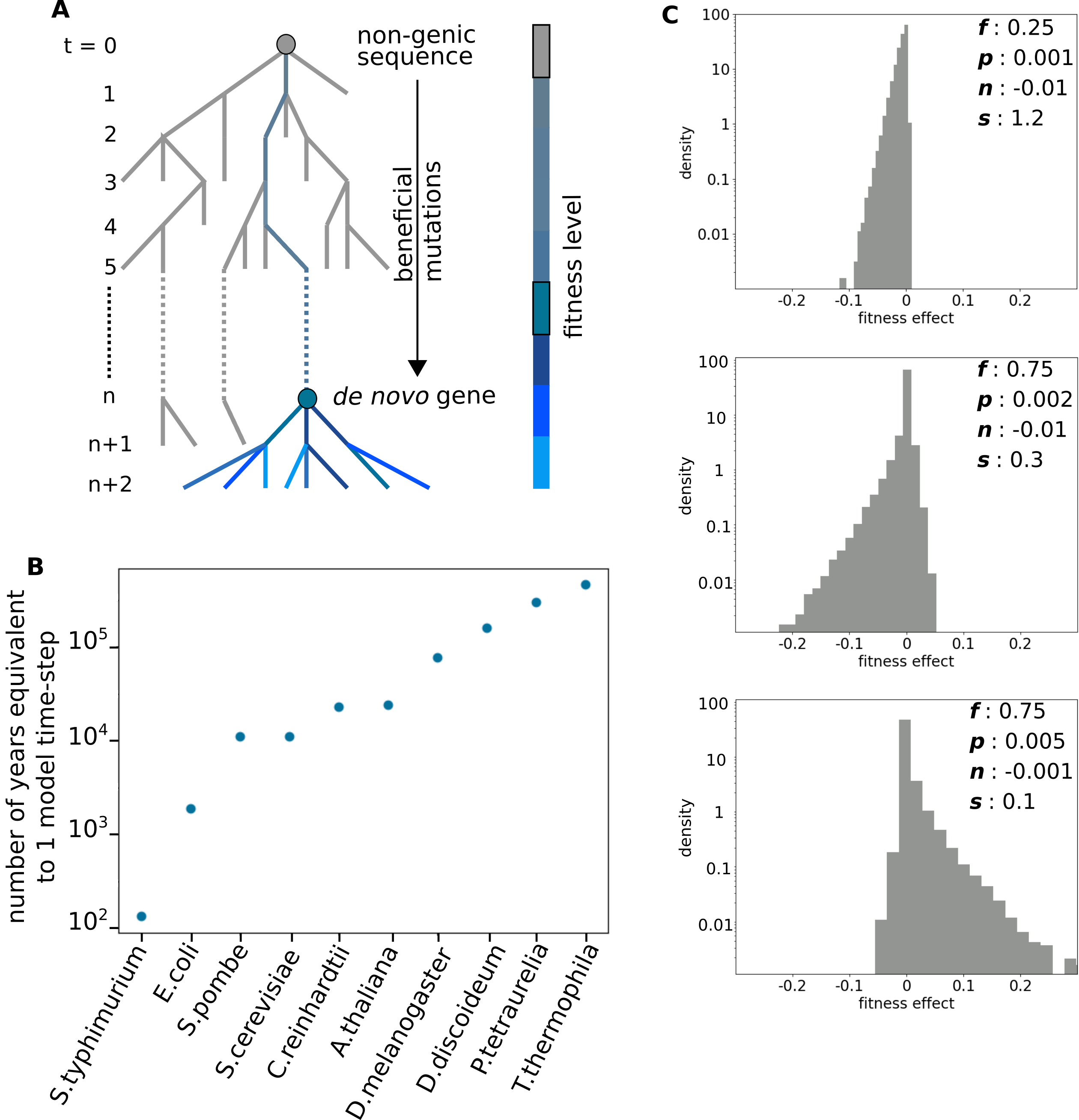}
\caption{\textbf{Time-scale and fitness effects of mutations in the model}. (A) Phylogenetic tree representing the evolution of the non-genic locus into a \denovo gene. Time steps $t$ count the generations in the model, which represent the average time for a mutation to occur in the locus. The grey dot at $t=0$ represents the initial non-genic sequence. Grey branches represent lineages that die out, and colored branches represent the lineage that gets fixes in the population. Fitness levels of colored branches in the fixed lineage are indicated in the color bar; the fitness level corresponding to non-genic sequences and to the genic threshold are outlined as the grey and blue box respectively. The blue dot at $t=n$ represents the most recent common ancestor of all surviving lineages whose fitness contribution is above the genic threshold. In the model, gene birth is said to have occurred at time step $t=n$. 
(B) Estimates of the number of years equivalent to a single time-step of the model in the different species listed on the x-axis. See FigS\ref{supptab:1} for calculations. 
(C) Distribution of fitness effects (DFE) for different model parameters. The top panel represents the DFE with the most deleterious and least beneficial mutations.  The bottom panel represents the DFE with the most beneficial and least deleterious mutations sampled in this work. The middle panel represents the DFE with parameters closest to those reported in~\cite{bondel2019inferring}. Values of parameters used to construct these DFEs are given alongside each histogram.}
\label{fig:schem}

\end{figure*}

Traditionally, the Wright-Fisher model is used to study the evolution of pre-existing genes; here we apply the model to non-genic loci and test the predictions it makes for \denovo gene birth. Initially, the distribution of $F_i$ is centered around $0$ and any small fitness effect is due to leaky expression. In each subsequent time-step $t+1$, the population is composed entirely of the offspring of individuals in the current time-step $t$ (Fig\ref{fig:schem}(A)). The probability that an individual leaves an offspring is proportional to the fitness $F_i$ of the locus, and individuals with $F_i \leq -1$ cannot produce offspring. 

Additionally, offspring in the model incur mutations. This sets the timescale of the model according to the mutation rate of the organism: a time-step in the model is roughly the time it takes for a mutation to occur in the locus. For a locus of ${\sim}100$ base pairs, a single model time-step can range between $\SIrange{100}{100000}{years}$ for different organisms (Fig\ref{fig:schem}(B), see also FigS\ref{suppfig:1}).

The fitness effects of mutations are drawn from the characteristic DFE for the locus (Fig\ref{fig:schem}(C)). Multiple studies indicate that long-tails are important features of DFEs, and the gamma distribution is a general form which can describe such long-tailed distributions ~\cite{eyre2007distribution}. Therefore, we choose to follow ~\cite{bondel2019inferring}, and represent DFEs as two-sided gamma distributions, and characterize them using four parameters: (i) average effect of beneficial mutations $p$, (ii) fraction of beneficial mutations $f$, (iii) average effect of deleterious mutations $n$, and (iv) the shape parameter $s$, where distributions with lower $s$ are more long-tailed. Note that, although experimental studies report DFEs across the whole genome, here we assume that the characteristics of the DFE of single loci are similar. The model describes the mutation types included in~\cite{bondel2019inferring}, which were single-nucleotide mutations and short indels (insertions or deletions of average length $\leq 10$ bp) \cite{ness2015extensive}. We account for differences in DFEs across species and locations on the genome by sampling across biologically reasonable values of these four parameters $p,f,n,s$ (see \nameref{methods:survey}). In all, we survey \num{225} parameter sets, and run \num{100} replicate populations for each set of DFE parameters.

We update populations for \num{2500} time-steps, equivalent to ~\numrange{0.2}{200} million years, depending on the organism and size of the locus (see \nameref{methods:pop_update}). In finite populations, we can trace the ancestry of each locus in each individual (see \nameref{methods:ancestry}), which allows us to track \textit{fixation events}: a mutant is said to have \textit{fixed} in the population if the ancestry of all individuals at some time-step $t$ can be traced back to a single individual at some previous time-step $t-t_{\rm fix}$. During the course of a simulation, populations undergo multiple fixation events. We say that \denovo gene birth occurs when the most recent mutant that gets fixed in the population is fitter than the predetermined \textit{genic} threshold (Fig\ref{fig:schem}(A)). For infinite populations, we model the evolution of the locus as a stochastic process (see \nameref{methods:inf_pop}). In this case, we say that gene birth occurs when the average fitness of the population is greater than the \textit{genic} threshold.

We verified that the model dynamics conform qualitatively with known results from the population genetics literature. For example, it is well-known that small populations are generally subject to stronger genetic drift, which makes it harder for fitter mutants to fix (\cite{gillespie2004population}, chapter 2). Consistently, in our simulations of populations of size N = \num{100}, \num{1000} and \num{5000}, the probability of fixation of fitter \denovo mutants that arise in smaller populations is lower (Table.\ref{tab1}(row1), FigS\ref{suppfig:0}(A), FigS\ref{suppfig:1}(A)). Additionally, mathematical models of the Moran process indicate that fitter mutants have a lower fixation probability and a shorter fixation time in smaller populations~\cite{diaz2014approximating}. In our model, across all fixation events, the mean fitness difference between the current fixed mutation and the previously fixed mutation tends to be much higher in larger populations (FigS\ref{suppfig:0}(B), FigS\ref{suppfig:1}(B)). And fixation occurs much faster in $N = \num{100}$ populations (Table.\ref{tab1}(row2), FigS\ref{suppfig:0}(C), FigS\ref{suppfig:1}(C) ).

\renewcommand\arraystretch{1.2}
\begin{table}[h!]
    \centering
        \caption{\textbf{Dynamics of fixation of beneficial mutations in finite populations.} (Row1) The fraction of systems in which a fitter mutant fixed in the population at least 75\% of the time across \num{2500} time-steps. (Row2) The average of mean-fixation time across \num{2500} time-steps.}
    \begin{tabular}[t]{m{0.37\linewidth} ccc}
        \toprule
        & N=100 & N=1000 & N=5000 \\
        \midrule
        Fraction of systems where a fitter mutant fixed at least \SI{75}{\percent} of the time & 0.57 & 0.70 & 0.77 \\
        Mean time-steps to fixation & 104.2 & 226.2 & 296.4 \\ 
        \bottomrule
    \end{tabular}

    \label{tab1}
\end{table}

In the following, we project these known results to the problem of gene birth, and derive expected bounds for the time scale and frequency of \denovo gene birth.

\section*{Predictions from the adaptive model}

\subsection*{Most of the genome is fertile ground}

\begin{figure*}

\includegraphics[width=0.95\linewidth]{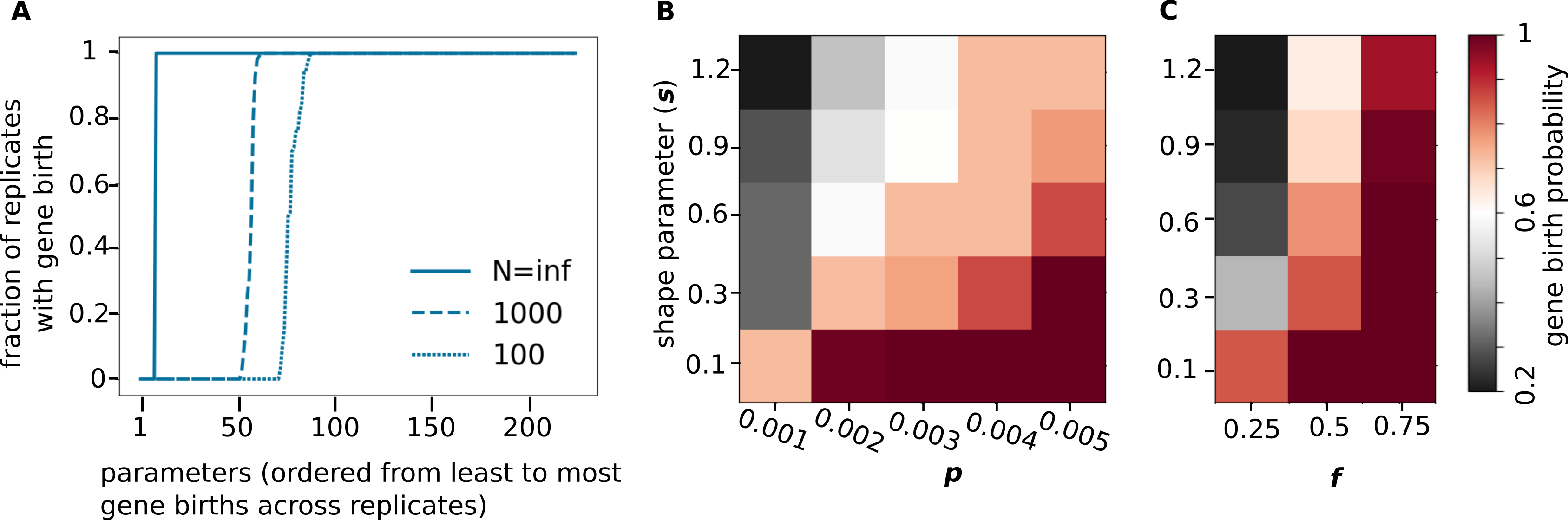}
\caption{ \textbf{Probability of gene birth.}  (A) The fraction of replicate systems at different parameter values that achieve gene birth. See also FigS\ref{suppfig:2}. (B,C) Trade-off between the shape parameter, $s$ (B) size of beneficial mutations, $p$ and (C) the frequency of beneficial mutations, $f$. We show results here for populations of size N = \num{1000}. See FigS\ref{suppfig:3} for N = 100, $\infty$. In the heatmaps, rows indicate values of the shape parameter and columns indicate values of parameters $p$ and $f$, respectively. Colors indicate the fractions of systems with gene birth as shown in the colorbar. See FigS\ref{suppfig:4} for effect of parameters on gene birth probability in populations of different sizes.}
\label{fig:2}

\end{figure*}

The parameters encoding different DFE represent the variety of different genomic regions across different species. We find that a majority of parameters in our survey are conducive to gene birth (Fig\ref{fig:2}(A), FigS\ref{suppfig:2}): across N = \num{100}, \num{1000} populations, \SI{60.9}{\percent} and \SI{72.4}{\percent}, respectively, of all parameters led to gene birth in all \num{100} replicate systems. Thus, our model suggests that gene birth due to spontaneous mutations should be a universal process.

Qualitatively, dynamics of gene birth in the model concur with results on fixation dynamics of beneficial mutations (Table.\ref{tab1}): gene birth is more prevalent in larger populations (Table.\ref{tab2} (row1)), and all parameters that allow gene birth in a small population also allow it in larger populations. Also, in finite populations, when gene birth occurred, it was faster in smaller populations (Table.\ref{tab2}(row2), Fig\ref{fig:3}, FigS\ref{suppfig:2}). We anticipate that the actual rate of gene birth in different organisms should scale according to features such as its generation time and mutation rate (Fig\ref{fig:schem}(B), FigS\ref{supptab:1}). 

\renewcommand\arraystretch{1.2}
\begin{table}[htbp!]
    \centering
        \caption{\textbf{Dynamics of gene birth for a genic threshold of 0.1.} (Row1) The fraction of parameters out of 225 in which gene birth was observed. For finite populations, this is the number of parameters where gene birth occurred in at least one out of 100 replicate populations. (Row2) Average time to gene birth.}
    \begin{tabular}{m{0.37\linewidth}ccc}
        \toprule
        & N=100 & N=1000 & N=$\infty$ \\
        \midrule
        Parameters with observed gene birth & 68.0\% & 76.9\% & 96.4\% \\
         Mean time-steps to gene birth & 380.7 & 452.4 & 357.7 \\ 
        \bottomrule
    \end{tabular}

    \label{tab2}
\end{table}

\subsection*{Rare, large beneficial mutations suffice for gene birth}

Reasonably, gene birth in the model was more likely when the frequency $f$ and average size $p$ of beneficial mutations are higher, and the size of deleterious mutations $n$ is lower (FigS\ref{suppfig:4}). In addition, we find that the shape parameter $s$ plays a crucial role: gene birth occurred even in populations with small values of parameters $f$ and $p$, provided the DFE of mutations is long-tailed (i.e., small values of $s$) (Fig\ref{fig:2}(B,C), see also FigS\ref{suppfig:3}). This suggests that large-effect beneficial mutations are sufficient for gene birth, even when they are rare.


\subsection*{Gene birth is practically inevitable under DFE fluctuations}
The DFE of a genomic locus is unlikely to remain the same over periods of millions of years that we simulate here. Broadly, there are three ways in which the DFE shifts in nature:
\begin{itemize}
    \item Over relatively short periods of time, and under fairly constant environments, organisms experience `diminishing returns epistasis', whereby the fitness gains due to beneficial mutations are smaller in relatively fit individuals than unfit individuals. Therefore, diminishing returns epistasis is likely to lead to decreased fitness gains along adaptive trajectories~\cite{wunsche2017diminishing}. In terms of the model, this would look like the DFE parameter $p$ reducing over time as fitter mutants undergo fixation in populations. 
    
    \item Over longer periods, environmental changes could lead to DFE variations due to changes in the magnitude of mutation effects, maybe even switching the sign of some mutations from beneficial to deleterious and vice-versa~\cite{das2020predictable}.
    
    \item The DFE can also change because of changes in frequencies of different types of mutations, such as transitions or transversions. This can happen when there is a shift in mutational biases, such as in mutator strains in \textit{E. coli}~\cite{sane2020shifts}.
\end{itemize}

Here, we test how fluctuations in the DFE could effect gene birth. We model DFE fluctuations by allowing one of the parameters of the DFE to shift to a neighbouring value at each time-step. For each initial parameter set, we run 10 replicate populations.

Now, the time-scale of each step in our model being of the order of hundreds to thousands of years, we do not expect the short-term effects of diminishing returns epistasis to persistently affect gene birth. Among the long-term fluctuations, we anticipate that environmental changes affect not only the DFE of new mutations, but also fitness contributions of preexisting genes. Such considerations are beyond the scope of the current model. Thus, the fluctuations we test here best represent shifts in mutational biases, which are expected to only affect the DFE of new mutations.

Studies indicate that shifts in mutational bias generally increase the rate of beneficial mutations~\cite{sane2020shifts}. But we allow fluctuations that decrease beneficial mutations; these potentially represent depletion of easily accessible beneficial mutations when the same mutational bias operates over long periods.

\begin{figure*}

\includegraphics[width=0.95\linewidth]{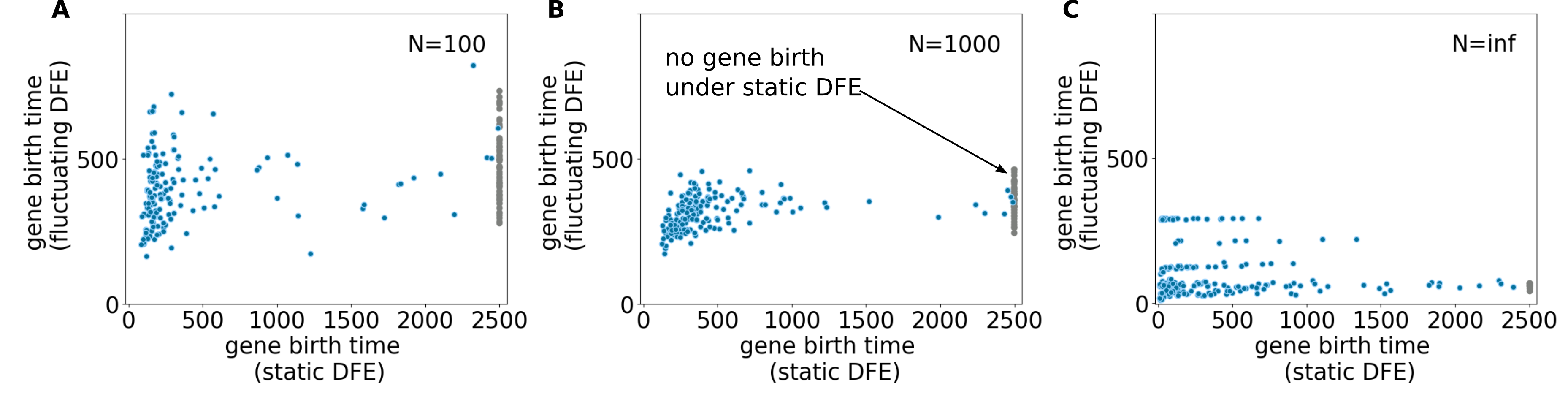}
\caption{ \textbf{Scatter plots comparing gene birth time in systems with static vs. fluctuating DFE.} For each point, the parameters for the static DFE and the initial parameters for fluctuating DFE are the same. The x-axis represents average time of gene birth across 100 replicate systems with static DFE, and the y-axis represents the average time of gene birth across 10 replicate systems when the DFE fluctuates, for (A) N = \num{100},  (B) N = \num{1000},  and (C) N = $\infty$. The grey points that accumulate at x = \num{2500} represent parameter values at which gene birth did not occur in static DFE systems. See also FigS\ref{suppfig:5},\ref{suppfig:6},\ref{suppfig:7}}
\label{fig:3}

\end{figure*}

We find that allowing DFE parameters to fluctuate makes gene birth almost inevitable in systems of all population sizes tested, and regardless of initial parameters. In Fig\ref{fig:3} (A,B,C), the grey points represent parameter values that did not lead to gene birth in any replicate population with static DFE. But under fluctuating DFE, gene birth occurs in all populations. For a large proportion of parameters, gene birth was faster, and gene birth times fall in a narrow range under fluctuating DFE (Fig\ref{fig:3}, see also FigS\ref{suppfig:5}).

To understand this high probability of gene birth, we notice that although each time-step involves only a small change in DFE parameters, this scheme of fluctuations uniformly samples all available DFEs. Over the \num{2500} time-steps of the simulation, irrespective of initial parameters, almost all parameters are visited at least once (FigS\ref{suppfig:6},\ref{suppfig:7}). Given that most DFEs are already highly conducive to gene birth (Fig\ref{fig:2}(A)), fluctuations in DFE allow all populations to spend substantial amounts of time in highly permissive parameters.

This in turn implies that gene birth depends on the duration a population spends in parameters conducive to gene birth, and is robust to the history of succession of these parameters. In other words, these systems are effectively equivalent irrespective of initial parameter values, thus also explaining the narrow distribution of time to gene birth. 

Altogether, our analysis of the evolutionarily plausible scenario where the DFE of loci is allowed to change over time bolsters the view that \denovo gene birth is a widespread process.

\section*{Discussion}

The study of \denovo genes draws attention to problems with our very conception of genes ~\cite{portin2017evolving}. It is increasingly apparent that organisms adapt not only by optimizing preexisting genetic modules to environmental parameters, but also through the invention of new genes, which exist transiently ~\cite{palmieri2014life}. How often genes are born \denovo, the time-scale of gene birth, and the time for which they persist, are therefore important questions that are fundamental to our understanding of adaptive evolution.

We use a simple mathematical model to specifically illustrate the process by which beneficial mutations accumulate at non-genic loci and turn them into genes. In our model, \denovo gene birth is a prevalent process, and particularly, rare, large-effect mutations play a major role in facilitating gene birth. Depending on the organism, a simulation of the model lasts for on the order of \numrange{e5}{e6} years, and gene birth is highly likely within this time frame for a large range of parameters. Thus, our results lead us to put forth the hypothesis that that \denovo gene birth is an unavoidable consequence of the constant adaptive evolution of non-genic loci.

Here, we use a coarse-grained definition of a gene, represented only by its fitness contribution. While this approach is immensely simple, it comes with trade-offs; firstly, birth of specific types of genes, such as anti-freeze proteins \cite{zhuang2019molecular} or miRNA \cite{lu2018possibility}, might display very different dynamics. Particularly, emergence of function in protein coding \denovo genes is very likely linked to the evolution of structural features~\cite{papadopoulos2021intergenic}. Whereas the model is agnostic to the molecular mechanisms by which functionality is achieved. 

On the other hand, this feature of the model is also useful, because in reality it is difficult to identify the emerging function of a \denovo gene \textit{a priori}, since we cannot envision what new factor in its environment the organism is next going to leverage. That is, new genes and their functions can only be identified with certainty post their establishment. But the evolutionary history of well-known \denovo genes~\cite{zhuang2019molecular}, and genomic studies that identify `proto-genes'~\cite{vakirlis2020novo} indicate that the fitness contribution of new genes is built up gradually. In this scenario, this framework allows a glimpse into the dynamics of gene birth irrespective of the exact function of the new gene.

Secondly, our model cannot capture non-adaptive aspects of genome evolution. For example, we cannot test here the role of pre-adaptation: a non-adaptive process that leads to sequences evolving away from harmful phenotypes, but does not prescribe whether and how sequences gain new functions~\cite{masel2006cryptic, wilson2017young}. 

Nevertheless, the present framework allows us to base our assumptions on multitudes of experiments that measure organismal fitness in terms of an easily accessible, universal quantity: the relative growth rate. Numerous studies measure the effect of mutations on organismal fitness in terms of the DFE, which allows us to investigate the contribution of spontaneous mutations towards gene birth. 

We use the DFE as a one-dimensional proxy for the various properties of genomic loci where \denovo gene birth is observed, such as high GC content, presence of bidirectional promoters, and presence of meiotic hotspots~\cite{vakirlis2018molecular}. For example, we imagine that at transcriptionally permissive genomic sequences, such as bidirectional promoters and nucleosome-free meiotic hotspots, any mutations are more likely to be expressed, and therefore are more likely to display DFE with higher values of $p$ and $n$ parameters. Moreover, the fitness contribution of any one locus necessarily depends on the cellular context through its interactions with other genes~\cite{wei2019patterns}. In this sense, the DFE also subsumes properties of the gene interaction network.


Ideally, a model of gene birth would describe the evolution of both expression level and functionality. In this sense, currently available DFE measurements are limited in that they do not disentangle the contributions to a mutation’s fitness effect due to changes in expression level versus changes in adaptive value. Such data would yield important insights into the dynamics of gene birth. For example, the birth of the \denovo gene poldi in the house mouse is attributed mostly to mutations in regulatory regions that led to increased expression \cite{heinen2009emergence}. More generally, the availability of such data could help resolve whether the prevalent mode of birth of protein coding \denovo genes is `expression first’, where loci are pervasively expressed and evolve functional features subsequently, or ‘ORF first’
\footnote{Open Reading Frames (ORFs) are genomic sequences encoding a sequence of amino acids uninterrupted by stop codons; all cellular proteins are expressed from ORFs.}, where loci that already possess ORFs gain expression~\cite{schlotterer2015genes}.

The present results provoke a natural question: why do we not see many more \denovo genes? While gene birth is likely to be a frequent and continuous process, the number of genes in a species is observed to remain fairly constant \cite{schlotterer2015genes}. One reason for this apparent discrepancy is that recently emerged genes, typically with low expression levels, are much more likely to die than older genes, that are consistently expressed \cite{lu2018possibility, palmieri2014life}. A second reason is the inability of current techniques to reliably detect \denovo genes: new genes differ from established genes in sequence properties, such as length, disordered regions etc. \cite{van2019novo}. Therefore, computational tools that are used to identify \denovo genes, which are based on sequence properties learnt from established genes, under-count new genes. Potentially, these discrepancies could be resolved by replacing \denovo gene identification methods that rely solely on sequence properties by methods that measure their fitness contributions. For example in \cite{vakirlis2020novo}, the fitness contributions of a chosen set of newly emerging genes are measured under conditions of gene disruption and over-expression. And in \cite{lu2018possibility}, the fitness effect of ablating \denovo genes was measured to examine how new genes die. We envisage that large-scale measurements that employ such techniques and similar approaches could shed light on the process of emergence of \denovo genes. 


\section*{Methods}

\subsection{Surveying the space of DFEs in populations of various sizes}\label{methods:survey}

We scan across DFEs with $p = [0.001, 0.002, 0.003, 0.004, 0.005]$, $f = [0.25, 0.5, 0.75]$, $n = [0.001, 0.005, 0.01]$ and $s = [0.1, 0.3, 0.6, 0.9, 1.2]$. And we look at populations of sizes $N = [100, 1000,  \infty]$. For each parameter set, we simulate 100 replicate systems with finite population sizes, and simulate once for the infinite populations. In all, we look at \num{45225} systems. We additionally look at fixation dynamics for \num{22500}  populations of size N = \num{5000} (100 replicates across all parameter values).
All codes used to generate and analyze data are written in Python3.6.

\subsection{Method to update population fitness for finite populations}\label{methods:pop_update}

For a population of size $N$, fitness of individuals at time-step $t$ are stored the vector $F_t \in \mathbb{R}^{N \times 1}$, where the fitness of some individual $i$ is $F_t(i)$.
Now, only individuals with fitness $> -1$ are viable, and capable of producing progeny. 

Individuals in the current population that produce progeny are chosen on the basis of their relative fitness.
Let $\minfit_t$ be the minimum fitness among viable individuals in $F_t$.

We define $\allfit_t = \sum_{j} \l(1 + F_t(j) - \minfit_t \r)$, for $j$ such that $F_t(j) >-1$.
The normalized relative fitness of individuals is then given by $\relfit_t \in [0,1]^{NX1}$, where
\begin{align*}
    & \textrm{relfit}_t(i) = \frac{1 + F_t(i) - \textrm{minfit}_t}{\allfit_t},  & \forall i \; \textrm{s.t.}\; F_t(i) > -1\\
    \textrm{and,~} & \relfit_t(i) = 0, & \forall i \; \textrm{s.t.} \; F_t(i) \leq -1~
\end{align*}

Let $\Anc_{t+1} \in \mathbb{N}^{NX1}$ be the list of individuals chosen from the current time-step $t$ to leave progeny. In other words, $\Anc_{t+1}$ is the list of ancestors of the population at time-step $t+1$.
We define
$\cumfit_t$ as the cumulative sum of $\relfit_t$, and
$\Anc_{t+1}(i) = \min \l( \l\{ j | \; \cumfit_t(j) \ge U(0,1) \r\} \r)$. 
Here $U(0,1)$ is a uniform random number between 0 and 1. This method ensures that the probability of choosing an individual is proportional to its fitness.

Progeny of the current population incur mutations. The values of fitness effects of mutations incurred by each individual at time-step $t$ is stored in $\mut_t \in \mathbb{R}^{N \times 1}$, where
\begin{align*}
    & \mut_t(i) = \Gamma\l(s, \frac{p}{s} \r) \iff \Ber(f) > 0~,\\
    \textrm{and,~}\; & \mut_t(i) = \Gamma\l(s, \frac{n}{s} \r) \iff \Ber(f) = 0~.    
\end{align*}
Here $\Gamma\l(\kappa,\theta \r)$ represents a number drawn from the gamma distribution with shape parameter $\kappa$ and scale parameter $\theta$, and $\Ber(p)$ is the Bernoulli random variable which equals 1 with probability $p$.
The updated fitness levels of the population is then given by
$F_{t+1}(i) = F_t\l( \Anc_{t+1}(i) \r) + \mut_t(i)$.

\subsection{Tracing ancestry and finding the fitness of the fixed mutation}\label{methods:ancestry}
In order to find the fitness value of the mutant fixed in the population at time-step $t$, we start with the list of ancestors of individuals $\Anc_t$ at time-step $t$. 
\newline Let $X_t = \{i, \forall i \in \Anc_t\}$ be the set of unique ancestor identities. We then recursively find $X_{t-n} = \{i, \forall i \in \{\Anc_{t-n}(j),\, \forall j \in X_{t-n+1}\}\}$ as the set of unique ancestor identities for $n = 1,2,3...t_0$, where $X_{t-t_0}$ is the first singleton set encountered. This set contains a single individual at time-step $t-t0-1$, whose mutations are inherited by every individual at time-step $t$. And the fitness value of the mutant fixed in the population at time-step $t$ is then  $F_{t-t_0-1}(i), \;\textrm{where}\; i \in X_{t-t0}$.
 
\subsection{Method to update population fitness for infinite populations}\label{methods:inf_pop}
In order to look at the evolution of fitness in infinite populations, we fix a reasonable bound, $F_{\max} >> \textrm{genic-threshold}$, on the maximum value of fitness, and use $F_{\min} = -F_{\max}$ as the minimum value of fitness. We also discretize the fitness values into levels $F^i$ separated by intervals of size $\delta$, such that $F^{i+1} - F^i = \delta$. 

Let $\mut(i\rightarrow j)$ be the probability density of a mutation of effect size $F^j - F^i$, where
 
\begin{align*}
    &\mut(i\rightarrow j) = f \cdot \Pr\l[\Gamma\l(\frac{p}{s},s\r)   = F^j - F^i \r], & \textrm{if}\; F^j > F^i\\
    &\mut(i\rightarrow j) = (1 - f) \cdot \Pr \l[\Gamma \l(\frac{n}{s},s\r)  = F^i - F^j \r], & \textrm{if}\; F^i>F^j
\end{align*}
The evolution of population fitness is then described by a transition matrix $T$, where
\begin{align*}
 & T(i,j) = \frac{\mut(i \rightarrow j)}{\sum_{k = - F_{\max}}^{F_{\max}} \mut(i \rightarrow k)}, & \; \textrm{for} \; F^i > -1\\
 \textrm{and,} & \qquad T(i,j) = 0, \qquad & \textrm{for} \; F^i \leq -1 
\end{align*}
Let the probability of occupancy of any fitness value $F^j$ at time-step $t$ be $P^j_t$. Here, relative fitness levels can be calculated as $\relfit^i = 1 - F^i - F_{\min}$. Then,
\begin{align*}
    P^j_{t+1} = \frac{\sum_{i=-F_{\max}}^{F_{\max}}\l(P^i_t \cdot \relfit^i \cdot T(i,j)\r)}{\sum_{i=-F_{\max}}^{F_{\max}}\l(P^i_t \cdot \relfit^i\r)}~. 
\end{align*}

\bibliography{genebirth_ref}

\begin{thebibliography}{10}
\providecommand \doibase [0]{http://dx.doi.org/}%

\bibitem{portin2017evolving}
Portin P, Wilkins A. The evolving definition of the term “gene”. {\it
  Genetics} 2017\string; 205(4)\string: 1353--1364.

\bibitem{tautz2011evolutionary}
Tautz D, Domazet-Lo{\v{s}}o T. The evolutionary origin of orphan genes. {\it
  Nature Reviews Genetics} 2011\string; 12(10)\string: 692--702.

\bibitem{vakirlis2020synteny}
Vakirlis N, Carvunis AR, McLysaght A. Synteny-based analyses indicate that
  sequence divergence is not the main source of orphan genes. {\it Elife}
  2020\string; 9\string: e53500.

\bibitem{vakirlis2018molecular}
Vakirlis N, Hebert AS, Opulente DA, et al. A molecular portrait of de novo
  genes in yeasts. {\it Molecular Biology and Evolution} 2018\string;
  35(3)\string: 631--645.

\bibitem{jensen2013dealing}
Jensen TH, Jacquier A, Libri D. Dealing with pervasive transcription. {\it
  Molecular cell} 2013\string; 52(4)\string: 473--484.

\bibitem{van2019novo}
Van~Oss SB, Carvunis AR. De novo gene birth. {\it PLoS genetics} 2019\string;
  15(5)\string: e1008160.

\bibitem{carvunis2012proto}
Carvunis AR, Rolland T, Wapinski I, et al. Proto-genes and de novo gene birth.
  {\it Nature} 2012\string; 487(7407)\string: 370--374.

\bibitem{hayashi2003can}
Hayashi Y, Sakata H, Makino Y, Urabe I, Yomo T. Can an arbitrary sequence
  evolve towards acquiring a biological function?. {\it Journal of molecular
  evolution} 2003\string; 56(2)\string: 162--168.

\bibitem{yona2018random}
Yona AH, Alm EJ, Gore J. Random sequences rapidly evolve into de novo
  promoters. {\it Nature communications} 2018\string; 9(1)\string: 1--10.

\bibitem{sane2020shifts}
Sane M, Diwan GD, Bhat BA, Wahl LM, Agashe D. Shifts in mutation spectra
  enhance access to beneficial mutations. {\it bioRxiv} 2020.

\bibitem{bondel2019inferring}
B{\"o}ndel KB, Kraemer SA, Samuels T, et al. Inferring the distribution of
  fitness effects of spontaneous mutations in Chlamydomonas reinhardtii. {\it
  PLoS biology} 2019\string; 17(6)\string: e3000192.

\bibitem{racimo2014approximation}
Racimo F, Schraiber JG. Approximation to the distribution of fitness effects
  across functional categories in human segregating polymorphisms. {\it PLoS
  Genet} 2014\string; 10(11)\string: e1004697.

\bibitem{huber2017determining}
Huber CD, Kim BY, Marsden CD, Lohmueller KE. Determining the factors driving
  selective effects of new nonsynonymous mutations. {\it Proceedings of the
  National Academy of Sciences} 2017\string; 114(17)\string: 4465--4470.

\bibitem{eyre2007distribution}
Eyre-Walker A, Keightley PD. The distribution of fitness effects of new
  mutations. {\it Nature Reviews Genetics} 2007\string; 8(8)\string: 610--618.

\bibitem{ness2015extensive}
Ness RW, Morgan AD, Vasanthakrishnan RB, Colegrave N, Keightley PD. Extensive
  de novo mutation rate variation between individuals and across the genome of
  Chlamydomonas reinhardtii. {\it Genome Research} 2015\string; 25(11)\string:
  1739--1749.

\bibitem{gillespie2004population}
Gillespie JH. {\it Population genetics: a concise guide}.
\newblock JHU press .
\newblock 2004.

\bibitem{diaz2014approximating}
D{\'\i}az J, Goldberg LA, Mertzios GB, Richerby D, Serna M, Spirakis PG.
  Approximating fixation probabilities in the generalized moran process. {\it
  Algorithmica} 2014\string; 69(1)\string: 78--91.

\bibitem{wunsche2017diminishing}
W{\"u}nsche A, Dinh DM, Satterwhite RS, Arenas CD, Stoebel DM, Cooper TF.
  Diminishing-returns epistasis decreases adaptability along an evolutionary
  trajectory. {\it Nature Ecology \& Evolution} 2017\string; 1(4)\string: 1--6.

\bibitem{das2020predictable}
Das SG, Direito SO, Waclaw B, Allen RJ, Krug J. Predictable properties of
  fitness landscapes induced by adaptational tradeoffs. {\it Elife}
  2020\string; 9\string: e55155.

\bibitem{palmieri2014life}
Palmieri N, Kosiol C, Schl{\"o}tterer C. The life cycle of Drosophila orphan
  genes. {\it elife} 2014\string; 3\string: e01311.

\bibitem{zhuang2019molecular}
Zhuang X, Yang C, Murphy KR, Cheng CHC. Molecular mechanism and history of
  non-sense to sense evolution of antifreeze glycoprotein gene in northern
  gadids. {\it Proceedings of the National Academy of Sciences} 2019\string;
  116(10)\string: 4400--4405.

\bibitem{lu2018possibility}
Lu GA, Zhao Y, Liufu Z, Wu CI. On the possibility of death of new
  genes--evidence from the deletion of de novo microRNAs. {\it BMC genomics}
  2018\string; 19(1)\string: 1--8.

\bibitem{papadopoulos2021intergenic}
Papadopoulos C, Callebaut I, Gelly JC, et al. Intergenic ORFs as elementary
  structural modules of de novo gene birth and protein evolution. {\it bioRxiv}
  2021.

\bibitem{vakirlis2020novo}
Vakirlis N, Acar O, Hsu B, et al. De novo emergence of adaptive membrane
  proteins from thymine-rich genomic sequences. {\it Nature communications}
  2020\string; 11(1)\string: 1--18.

\bibitem{masel2006cryptic}
Masel J. Cryptic genetic variation is enriched for potential adaptations. {\it
  Genetics} 2006\string; 172(3)\string: 1985--1991.

\bibitem{wilson2017young}
Wilson BA, Foy SG, Neme R, Masel J. Young genes are highly disordered as
  predicted by the preadaptation hypothesis of de novo gene birth. {\it Nature
  ecology \& evolution} 2017\string; 1(6)\string: 1--6.

\bibitem{wei2019patterns}
Wei X, Zhang J. Patterns and mechanisms of diminishing returns from beneficial
  mutations. {\it Molecular biology and evolution} 2019\string; 36(5)\string:
  1008--1021.

\bibitem{heinen2009emergence}
Heinen TJ, Staubach F, H{\"a}ming D, Tautz D. Emergence of a new gene from an
  intergenic region. {\it Current biology} 2009\string; 19(18)\string:
  1527--1531.

\bibitem{schlotterer2015genes}
Schl{\"o}tterer C. Genes from scratch--the evolutionary fate of de novo genes.
  {\it Trends in Genetics} 2015\string; 31(4)\string: 215--219.

\bibitem{farlow2015spontaneous}
Farlow A, Long H, Arnoux S, et al. The spontaneous mutation rate in the fission
  yeast Schizosaccharomyces pombe. {\it Genetics} 2015\string; 201(2)\string:
  737--744.

\bibitem{keightley2009analysis}
Keightley PD, Trivedi U, Thomson M, Oliver F, Kumar S, Blaxter ML. Analysis of
  the genome sequences of three Drosophila melanogaster spontaneous mutation
  accumulation lines. {\it Genome research} 2009\string; 19(7)\string:
  1195--1201.

\bibitem{lee2012rate}
Lee H, Popodi E, Tang H, Foster PL. Rate and molecular spectrum of spontaneous
  mutations in the bacterium Escherichia coli as determined by whole-genome
  sequencing. {\it Proceedings of the National Academy of Sciences}
  2012\string; 109(41)\string: E2774--E2783.

\bibitem{lind2008whole}
Lind PA, Andersson DI. Whole-genome mutational biases in bacteria. {\it
  Proceedings of the National Academy of Sciences} 2008\string; 105(46)\string:
  17878--17883.

\bibitem{long2016low}
Long H, Winter DJ, Chang AYC, et al. Low base-substitution mutation rate in the
  germline genome of the ciliate Tetrahymena thermophila. {\it Genome biology
  and evolution} 2016\string; 8(12)\string: 3629--3639.

\bibitem{ness2012estimate}
Ness RW, Morgan AD, Colegrave N, Keightley PD. Estimate of the spontaneous
  mutation rate in Chlamydomonas reinhardtii. {\it Genetics} 2012\string;
  192(4)\string: 1447--1454.

\bibitem{ossowski2010rate}
Ossowski S, Schneeberger K, Lucas-Lled{\'o} JI, et al. The rate and molecular
  spectrum of spontaneous mutations in Arabidopsis thaliana. {\it science}
  2010\string; 327(5961)\string: 92--94.

\bibitem{saxer2012whole}
Saxer G, Havlak P, Fox SA, et al. Whole genome sequencing of mutation
  accumulation lines reveals a low mutation rate in the social amoeba
  Dictyostelium discoideum. {\it Plos One} 2012.

\bibitem{sung2012extraordinary}
Sung W, Tucker AE, Doak TG, Choi E, Thomas WK, Lynch M. Extraordinary genome
  stability in the ciliate Paramecium tetraurelia. {\it Proceedings of the
  National Academy of Sciences} 2012\string; 109(47)\string: 19339--19344.

\bibitem{zhu2014precise}
Zhu YO, Siegal ML, Hall DW, Petrov DA. Precise estimates of mutation rate and
  spectrum in yeast. {\it Proceedings of the National Academy of Sciences}
  2014\string; 111(22)\string: E2310--E2318.

\bibitem{cassidy2012tetrahymena}
Cassidy-Hanley DM. Tetrahymena in the laboratory: strain resources, methods for
  culture, maintenance, and storage. {\it Methods in cell biology} 2012\string;
  109\string: 237--276.

\bibitem{milo2010bionumbers}
Milo R, Jorgensen P, Moran U, Weber G, Springer M. BioNumbers—the database of
  key numbers in molecular and cell biology. {\it Nucleic acids research}
  2010\string; 38(suppl\_1)\string: D750--D753.

\bibitem{petersen2016growth}
Petersen J, Russell P. Growth and the environment of Schizosaccharomyces pombe.
  {\it Cold Spring Harbor Protocols} 2016\string; 2016(3)\string:
  pdb--top079764.

\bibitem{harris2009chlamydomonas}
Harris EH. {\it The Chlamydomonas Sourcebook: Introduction to Chlamydomonas and
  Its Laboratory Use: Volume 1}. 1.
\newblock Academic press .
\newblock 2009.

\bibitem{ishida2017improved}
Ishida M, Hori M. Improved isolation method to establish axenic strains of
  Paramecium. {\it Japanese Journal of Protozoology} 2017\string;
  50(1-2)\string: 1--14.

\bibitem{fey2007protocols}
Fey P, Kowal AS, Gaudet P, Pilcher KE, Chisholm RL. Protocols for growth and
  development of Dictyostelium discoideum. {\it Nature protocols} 2007\string;
  2(6)\string: 1307--1316.

\bibitem{silva2009validation}
Silva RR, Moraes CA, Bessan J, Vanetti MCD. Validation of a predictive model
  describing growth of Salmonella in enteral feeds. {\it Brazilian Journal of
  Microbiology} 2009\string; 40(1)\string: 149--154.

\bibitem{fernandez2007drosophila}
Fern{\'a}ndez-Moreno MA, Farr CL, Kaguni LS, Garesse R. Drosophila melanogaster
  as a model system to study mitochondrial biology. In: Springer.  2007 (pp.
  33--49).

\bibitem{koornneef2001arabidopsis}
Koornneef M, Scheres B. Arabidopsis thaliana as an experimental organism. {\it
  e LS} 2001.

\end{thebibliography}

\section*{Acknowledgements}
This work was funded by the Institute for Basic Science, Grant IBS-R020.

\section*{Supplementary information}\label{supp}

\begin{FigS*}[h!]

\begin{center}
\includegraphics[width=0.5\linewidth]{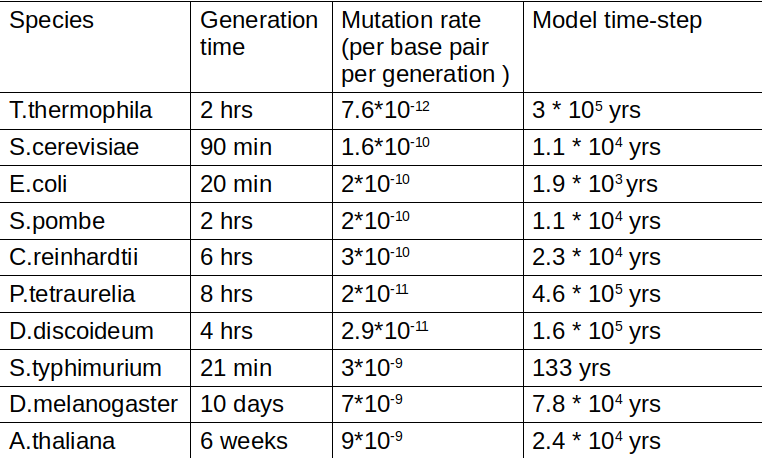}
\caption{Estimates of one model time-step in various species. The following references were used to obtain mutation rates: \cite{farlow2015spontaneous, keightley2009analysis, lee2012rate, lind2008whole, long2016low, ness2012estimate, ossowski2010rate, saxer2012whole, sung2012extraordinary, zhu2014precise}. And the following references were used for generation times: \cite{cassidy2012tetrahymena, milo2010bionumbers, petersen2016growth, harris2009chlamydomonas, ishida2017improved, fey2007protocols, silva2009validation, fernandez2007drosophila, koornneef2001arabidopsis}. Most references give a range of values or multiple values of mutation rates and generation times depending on the culture conditions used. In such cases, we picked the average reported value at a single culture condition that is consistent between the studies reporting mutation rate and generation time in any one organism. The model time-step calculated in column 3 is the time it takes for 1 mutation to occur in a locus of 100 bp.}
\label{supptab:1}    
\end{center}

\end{FigS*}


\begin{FigS*}

\includegraphics[width=1\linewidth]{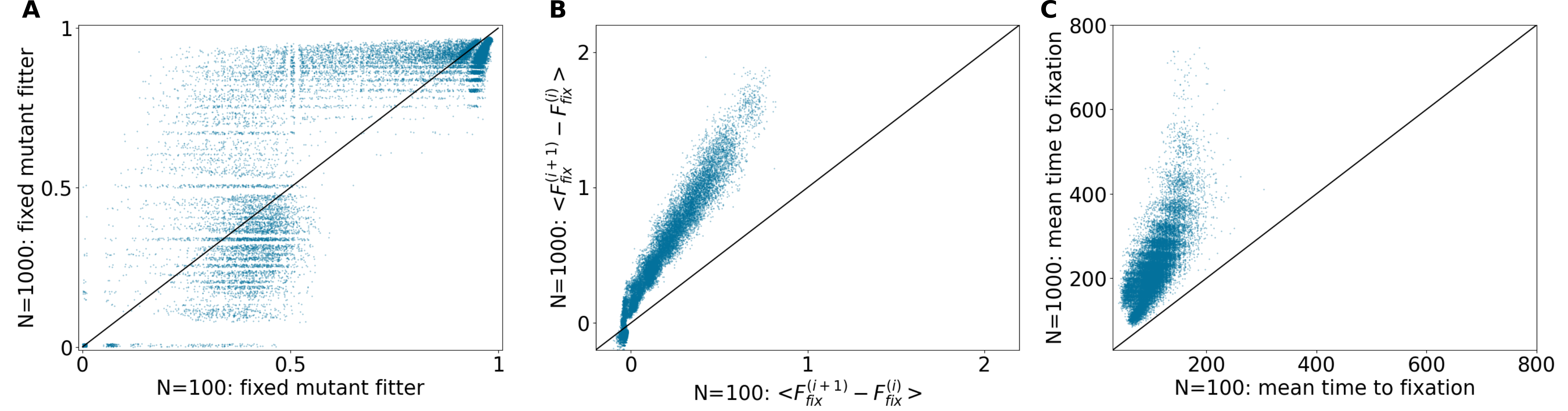}
\caption{ \textbf{Scatter plots comparing fixation dynamics in $N = \num{100}$ and \num{1000} populations.} For each population size, \num{22500} systems were analysed across all parameter values. For each point, the x-coordinate represents an N = \num{100} and the y-coordinate represents an N= \num{1000} population that have the same parameter values. (A) Fraction of times fixation of fitter mutant occurred, (B) mean fitness difference between current and previous fixed mutant, $\langle \Delta F_{\rm fix} \rangle = \langle F_{\rm fix}^{i+1} - F_{\rm fix}^i \rangle$, where $F_{\rm fix}^i$ is the fitness of the $i^{\rm th}$ mutant that fixed in the population, and (C) mean fixation time across \num{2500} time steps.} 
\label{suppfig:0}

\end{FigS*}

\begin{FigS*}[h!]

\hspace*{-1cm}
\includegraphics[width=1\linewidth]{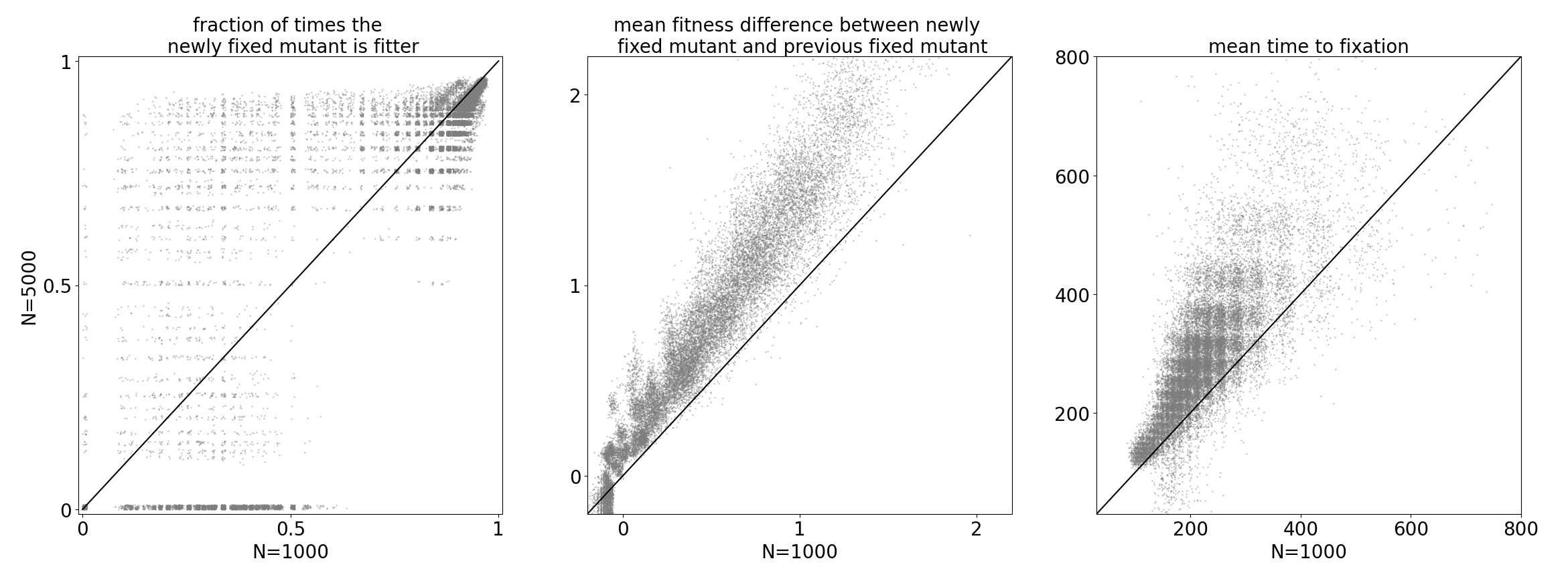}
\caption{ Scatter plots comparing fixation dynamics in N = 1000 and 5000 populations. For each population size, 22,500 systems were analysed across all parameter values. The x and y value for each point represents (A) Fraction of times fixation of fitter mutant occurred, (B) mean fitness difference between current and previous fixed mutant, (C) Mean fixation time across 2500 time steps, for an N = 1000 and an N = 5000 systems that have equal parameter values}
\label{suppfig:1}
\hspace*{-1.5cm}

\end{FigS*}


\begin{FigS*}[h!]

\includegraphics[width=0.95\linewidth]{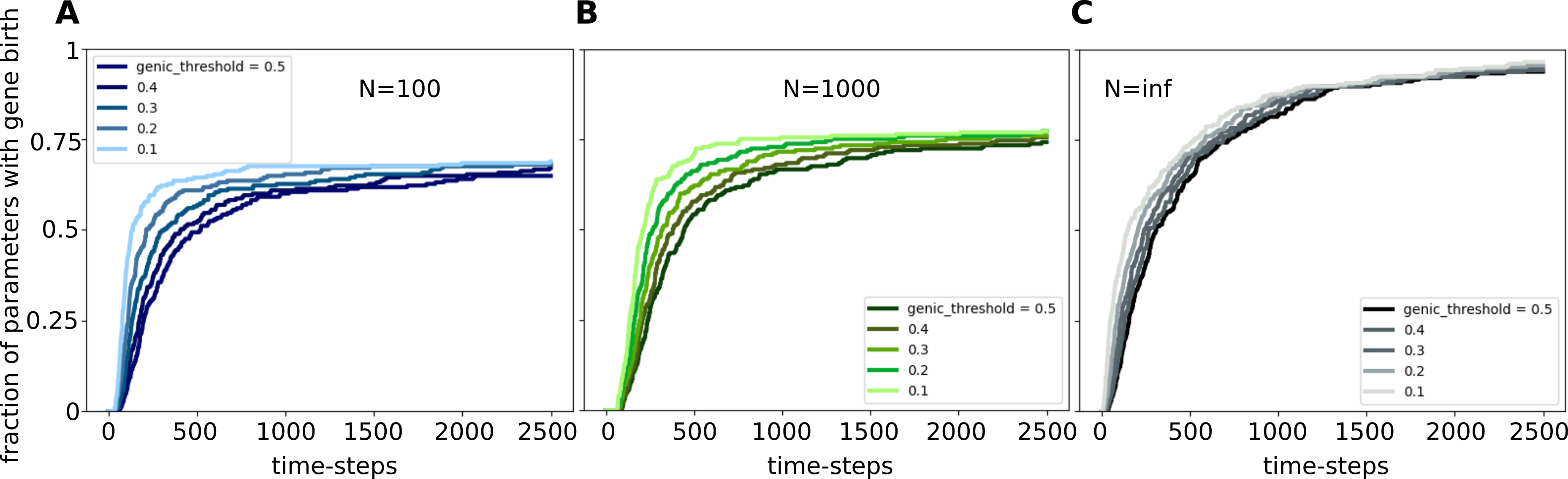}
\caption{ In (A,B,C) the x-axis represents time-steps and the y-axis represents the fraction of parameters tested for which gene birth occurred at least in 1 of the 100 replicate populations tested. Population sizes: (A) N=100, (B) N=1000, (C) infinite population.}
\label{suppfig:2}
\hspace*{-1cm}

\end{FigS*}

\begin{FigS*}[h!]

\begin{center}
\hspace*{-1cm}
\includegraphics[width=0.55\linewidth]{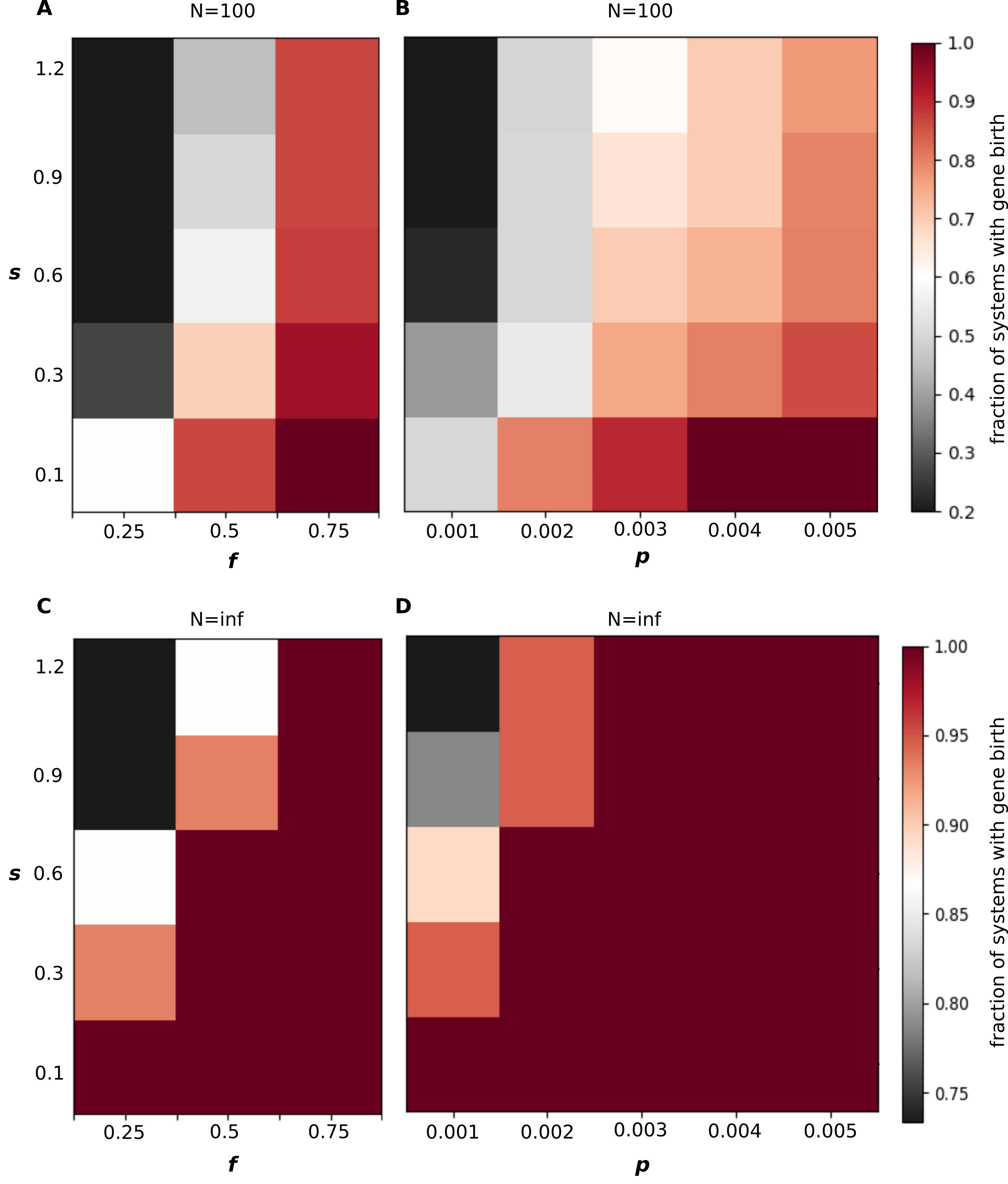}
\caption{Trade-off between the shape parameter and frequency and size of beneficial mutations. We show results here for populations of sizes (A,B) N = 100 and (C,D) N = inf. In the heatmaps of rows indicate values of the shape parameter, columns indicate values of (A,C) fraction of beneficial mutation \textit{f}, (B,D) mean size of beneficial mutations \textit{p}. Colors indicate the fractions of systems with gene birth as indicated by the colorbar.}
\label{suppfig:3}    
\hspace*{-1cm}
\end{center}

\end{FigS*}

\begin{FigS*}[h!]

\begin{center}
\hspace*{-1.5cm}
\includegraphics[width=\dimexpr\textwidth+3cm\relax]{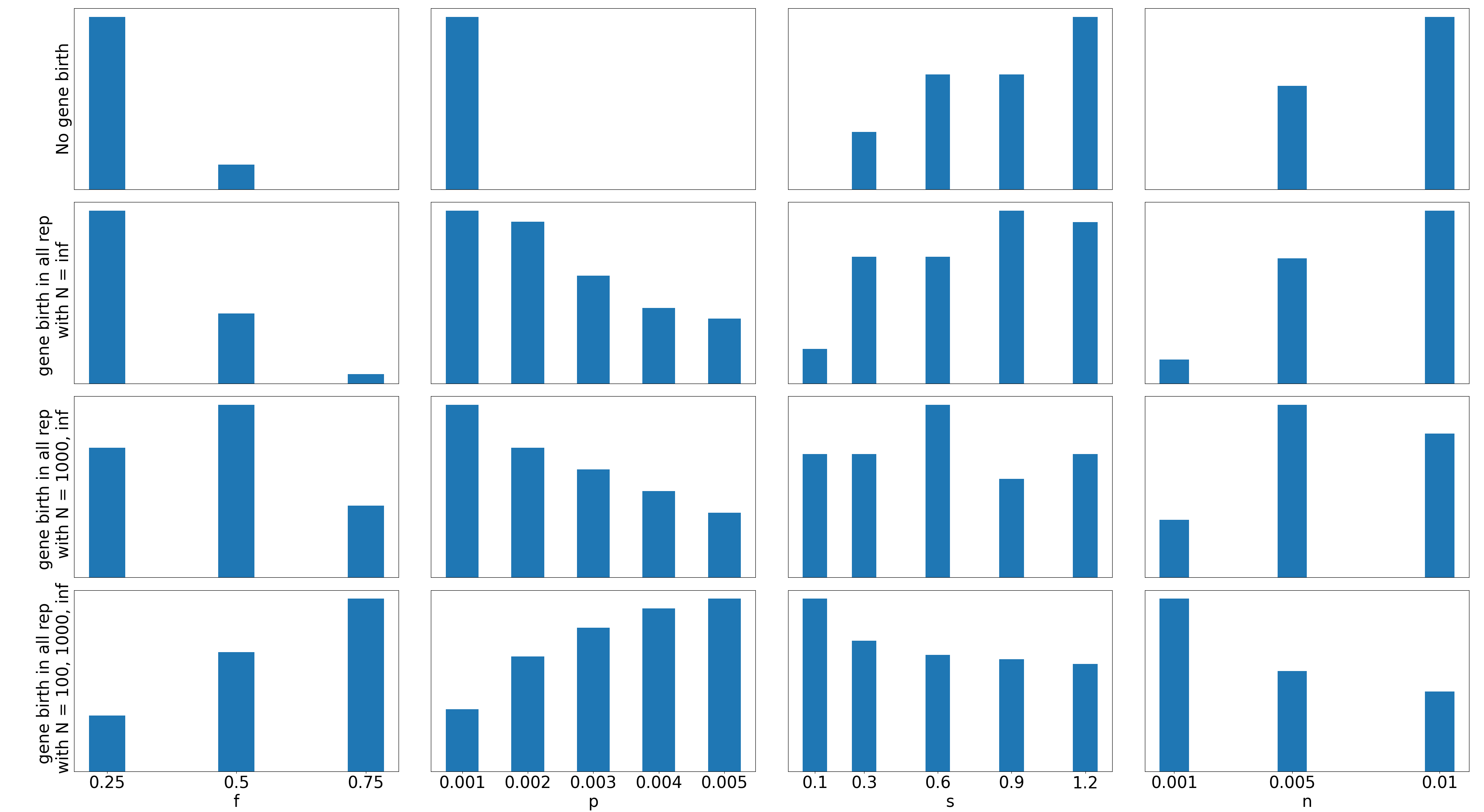}
\caption{Effect of model parameters on gene birth probability. Each row contains histograms for parameter values at which \textit{first row}: no gene birth occurred in any population, \textit{second row}: gene birth occurred in all 100 replicates only in N = inf populations, \textit{third row}: gene birth occurred in all 100 replicates only in N = 1000 and N = inf populations, \textit{fourth row}: gene birth occurred in all 100 replicates across all N = 100, 1000, inf. Columns correspond to different parameters, first column: \textit{f}, second column: \textit{p}, third column: \textit{s}, fourth column: \textit{n}}
\label{suppfig:4}
\hspace*{-2cm}
\end{center}

\end{FigS*}


\begin{FigS*}[h!]

\includegraphics[width=0.95\linewidth]{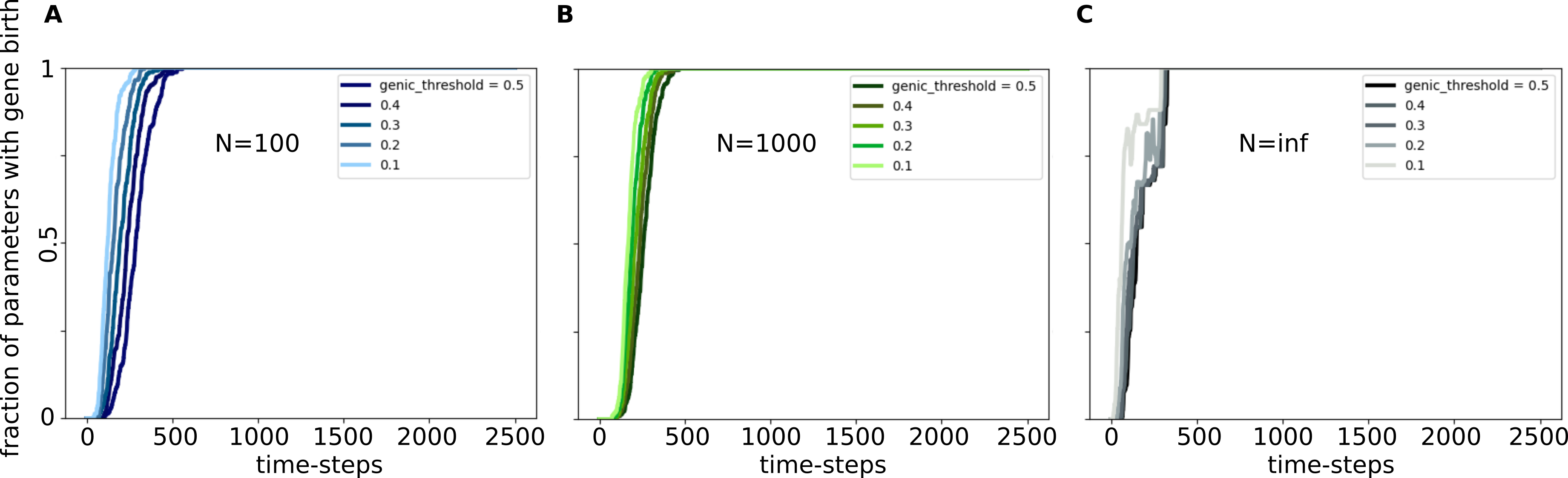}
\caption{ All parameters lead to gene birth within 2500 time-steps under the fluctuating DFE regime. In (A,B,C) the x-axis represents time-steps and the y-axis represents the fraction of parameters tested for which gene birth occurred at least in 1 of the 100 replicate populations tested. Population sizes: (A) N=100, (B) N=1000, (C) infinite population.}
\label{suppfig:5}
\hspace*{-1cm}

\end{FigS*}

\begin{FigS*}[h!]

\includegraphics[width=0.9\linewidth]{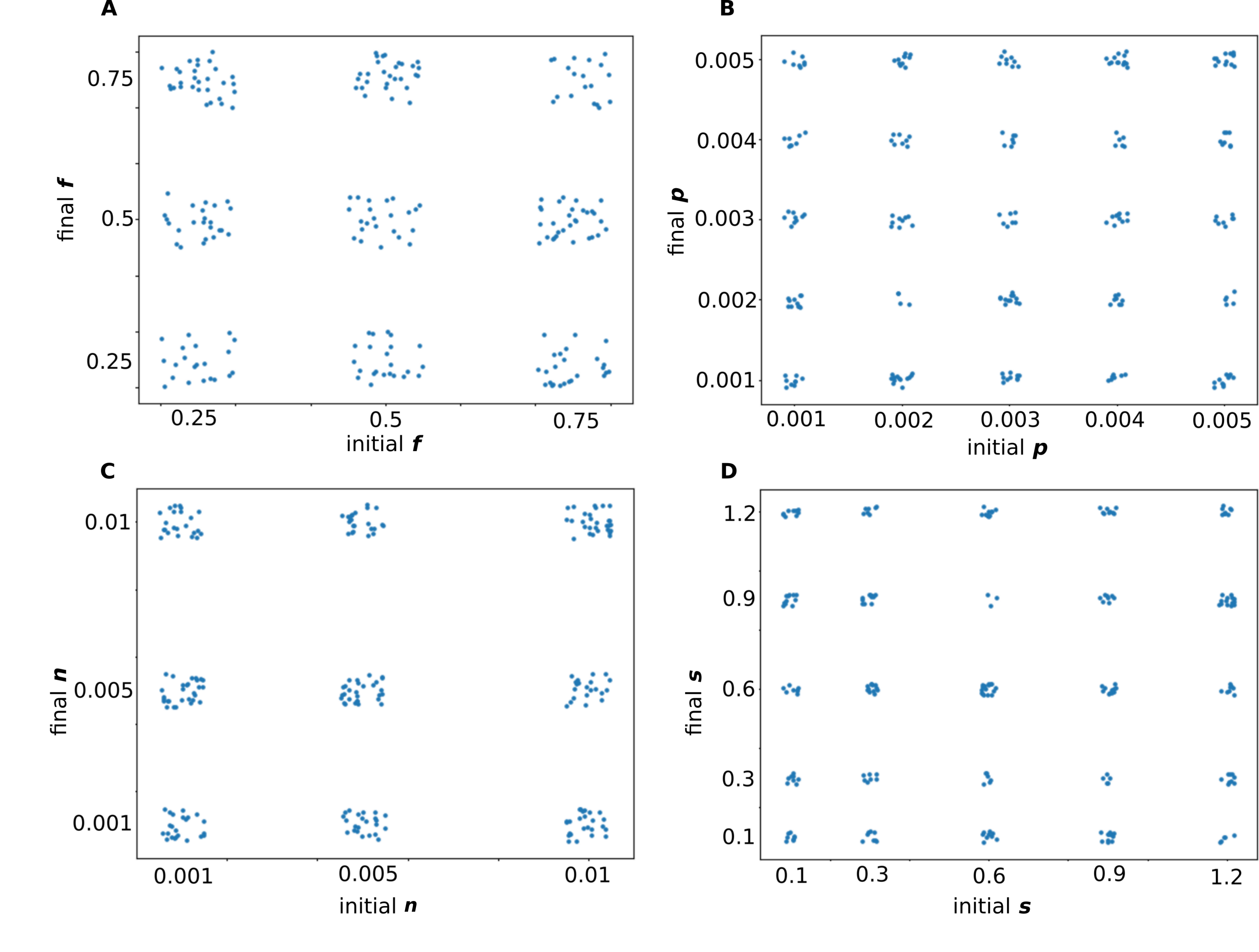}
\caption{Scatter plots for initial and final values of parameters under fluctuating DFE regime. Each point represents a distinct N = 1000 population. 2250 systems were analysed for this figure. Noise has been added to make the density of points more apparent. (A) fraction of beneficial mutations \textit{f}, (B) mean size of beneficial mutations \textit{p}, (C) mean size of deleterious mutations \textit{n}, (D) shape parameter \textit{s}}
\label{suppfig:6}

\end{FigS*}

\begin{FigS*}[h!]

\begin{center}
\includegraphics[width=0.75\linewidth]{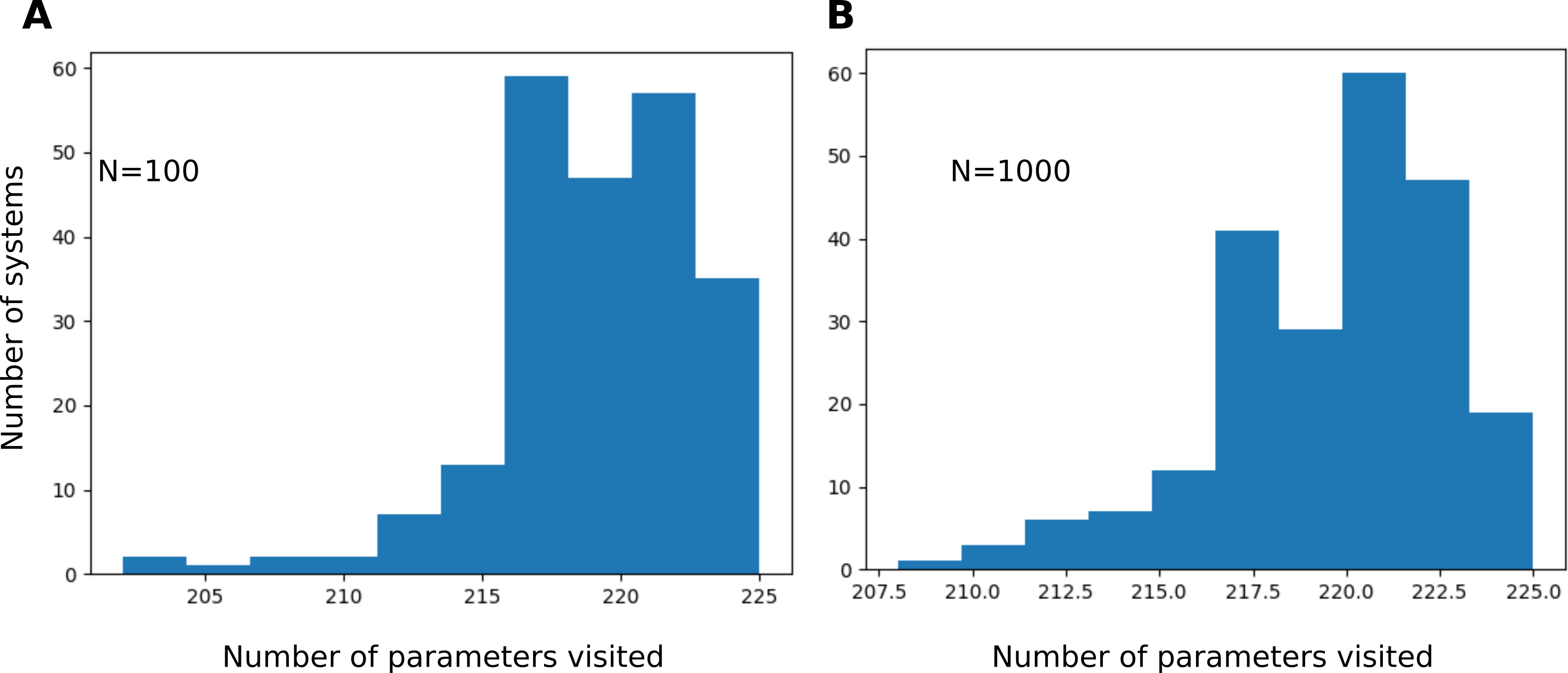}
\caption{Histograms for number of distinct parameters visited by populations under the fluctuating DFE regime. 2250 systems were analysed for each histogram. (A) N = 100, (B) N = 1000}
\label{suppfig:7}    
\end{center}

\end{FigS*}

\end{document}